\documentstyle[amssymb,aps,epsfig,float,twocolumn]{revtex}
\restylefloat{figure}

\newcommand{\cgo}{CuGeO$_3$}
\newcommand{\sds}[2]{{{\bf S}_{#1}\!\cdot\!{\bf S}_{#2}}}

\title{Quantum lattice fluctuations
  in a frustrated Heisenberg spin-Peierls chain\thanks{
    Dedicated to H.~B\"uttner on the occasion of his 60th birthday.}}
\author{A.~Wei{\ss}e$^a$, G.~Wellein$^b$ and H.~Fehske$^c$}
\address{$^a$Physikalisches Institut, Universit\"a{}t Bayreuth, 
  95440 Bayreuth, Germany\\
  $^b$Regionales Rechenzentrum Erlangen, Universit\"a{}t Erlangen, 
  91058 Erlangen, Germany\\
  $^c$Theoretische Physik, Universit\"a{}t des Saarlandes, 
  66041 Saarbr\"ucken, Germany\\
  {\rm (\today)}
  }
\address{~\parbox{14cm}{\rm
    \medskip
    As a simple model for spin-Peierls systems we study
    a frustrated Heisenberg chain coupled to optical phonons. 
    In view of the anorganic spin-Peierls compound \cgo{} we 
    consider two different mechanisms of spin-phonon coupling.
    Combining variational concepts in the adiabatic regime and
    perturbation theory in the anti-adiabatic regime we
    derive effective spin Hamiltonians which cover the dynamical
    effect of phonons in an approximate way. Ground-state phase
    diagrams of these models are determined, and the effect of frustration 
    is discussed. Comparing the properties of the ground state and
    of low-lying excitations with exact diagonalization data 
    for the full quantum spin phonon models, good agreement is found
    especially in the anti-adiabatic regime.
    }}

\begin{document}

\maketitle

\section{Introduction}

The effect of a Peierls instability in quasi-one-dimensional spin
systems, i.e. the instability of a uniform spin chain towards 
dimerization induced by the interaction with lattice degrees
of freedom, has attracted considerable attention over the last
decades. Starting in the seventies with organic compounds of
the TTF and TCNQ family~\cite{Mil83}, the interest in the Peierls 
instability was renewed with the discovery of a spin-Peierls (SP)
transition in the anorganic compound \cgo{} in 1993 by 
Hase~et~al.~\cite{HTU93}. The most significant feature distinguishing
\cgo{} from other SP-compounds is the high energy of the involved 
optical phonons, which is comparable to the magnetic exchange 
integral $J$. In contrast to the organic materials no softening of
these phonon modes is observed near the transition. Therefore 
the adiabatic treatment of the phonon subsystem used in the works of 
Pytte~\cite{Pyt74} or  Cross and Fisher~\cite{CF79},
does not seem appropriate to describe the SP transition in \cgo{},
although there are recent efforts in this direction~\cite{GW98}.
Rather one has to take into account the effect of quantum lattice 
fluctuations which tend to decrease the SP transition temperature respectively
the energy gap between the ground state and lowest excitations in the
dimerized phase. Unfortunately there are practically no analytic
methods to handle coupled systems of spins (electrons) and phonons
when all energy scales and coupling strengths are of the 
same order of magnitude. This is why many studies involving 
dynamical phonons rely on numerical methods, such as 
exact diagonalization~(ED)~\cite{WFK98,BFK98},
density matrix renormalization group~(DMRG)~\cite{CM96,BMH98} or 
Monte Carlo simulation~(MC)~\cite{MC}.
Only recently Zheng~\cite{Zhe97} developed an analytical approach to 
describe the SP instability of a XY spin chain, which is based on the 
unitary transformation method. It works well in the adiabatic and in 
the anti-adiabatic regime. In the latter case there are also some 
approaches to the Heisenberg spin chain interacting with optical phonons: 
Kuboki and Fukuyama~\cite{KF87} used perturbation theory to derive an 
effective spin Hamiltonian, while Uhrig~\cite{Uhr98} applied the 
flow-equation method~\cite{Weg94}.

As a simple model which contains all important features of a
SP system in the following we consider an antiferromagnetic Heisenberg 
chain coupled to a set of Einstein oscillators,
\begin{equation}
  H = H_{\rm s} + H_{\rm p} + H_{\rm sp}\label{model}
\end{equation}
with
\begin{eqnarray}
  H_{\rm s} & = & J \sum_i (\sds{i}{i+1} + \alpha\sds{i}{i+2})\,,\\
  H_{\rm p} & = & \omega_0 \sum_i b_i^+ b_i\,.
\end{eqnarray}
The interaction of spins and phonons, $H_{\rm sp}$, can be modelled in two
different ways,
\begin{eqnarray}
  H_{\rm sp}^{\rm loc} & = & \bar g \sum_i (b_i^+ + b_i)\sds{i}{i+1}\,,\\
  H_{\rm sp}^{\rm diff} & = & \bar g \sum_i (b_i^+ + b_i)
  (\sds{i}{i+1} - \sds{i}{i-1})\,,
\end{eqnarray}
where ${\bf S}_{i}$ denote spin-$1\over 2$ operators at lattice site
$i$, while $b_i^{+}$ and $b_i$ are phonon creation and annihilation 
operators, respectively. 
$H_{\rm sp}^{\rm loc}$ and $H_{\rm sp}^{\rm diff}$ differ in the 
mechanism, how the lattice influences the exchange integral. 
For $H_{\rm sp}^{\rm loc}$, the {\it local} coupling, one can think 
of a single harmonic degree of freedom directly modifying the magnetic 
interaction. In the context of \cgo{} this could correspond to 
side group effects (by the Germanium atoms) as discussed in Ref.~\cite{GKM}.
In the case of $H_{\rm sp}^{\rm diff}$, the {\it difference} coupling, 
the exchange depends directly on the spatial distance between 
neighbouring spins. 
Note that it is not possible to uniformly decrease or increase all 
exchange integrals with this type of spin phonon interaction.

Although $H_{\rm sp}^{\rm loc}$ seems to be more appropriate for \cgo{} 
we will consider both variants and compare their properties.
In addition we take into account a frustrating next-nearest 
neighbour interaction $J\alpha$, which in view of \cgo{} was introduced to
explain susceptibility data~\cite{CCE95}. As we will see below, the 
spin-phonon interaction is able to induce this kind of long ranged 
exchange as well.

Motivated by the success of methods combining unitary transformations 
with variational and numerical techniques, which we used to study the 
Peierls transition in the Holstein model of spinless fermions~\cite{WF98}, 
and inspired by the work of Zheng~\cite{Zhe97}, in this article we  
analyse the model~(\ref{model}) within the same framework.
In particular we focus on the ground-state phase diagram as a function
of spin-phonon coupling, frustration and phonon frequency, and
compare our results with exact diagonalization data.

\section{Effective spin models}

To describe a static lattice dimerization in the adiabatic case
of small phonon frequency $\omega_0$ we start with a unitary
transformation of $H$ which shifts the equilibrium position of
each oscillator by a constant amount alternating from site to site,
$\tilde H = \exp(S_1) H \exp(-S_1)$ with
\begin{equation}
  S_1 = {\Delta_\pi\over 2\bar g} \sum_i (-1)^i (b_i^+ - b_i)\,.
\end{equation}
For the terms involving phonons this yields 
\begin{eqnarray}
  \tilde H_{\rm p} & = & H_{\rm p} 
  - \omega_0 {\Delta_\pi\over 2\bar g}\sum_i (-1)^i (b_i^+ + b_i)\nonumber\\
  & & + N \omega_0 \left({\Delta_\pi \over 2\bar g}\right)^2\,,\\ 
  \tilde H_{\rm sp}^{\rm loc} & = & H_{\rm sp}^{\rm loc}
  -\Delta_\pi \sum_i (-1)^i \sds{i}{i+1}\,,\\
  \tilde H_{\rm sp}^{\rm diff} & = & H_{\rm sp}^{\rm diff}
  - 2 \Delta_\pi \sum_i (-1)^i \sds{i}{i+1}\,.
\end{eqnarray}
$\Delta_\pi$ will act as the variational parameter
describing the dimerization of the system.

\begin{figure}[htb]
  \epsfig{file= 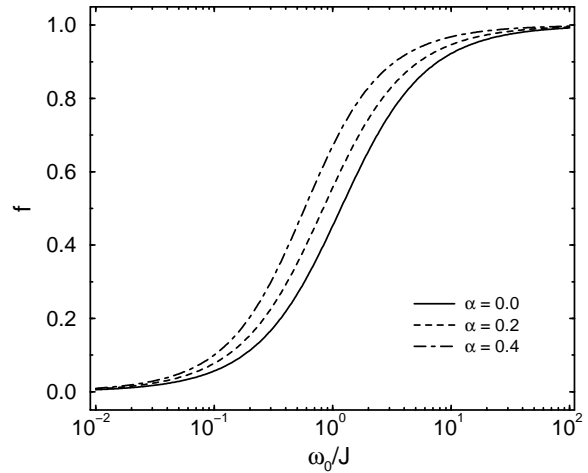, width = \linewidth}
  \caption{Variation of $f$ in the case of local coupling and
    lattice size $N=16$.}\label{figfopt}
\end{figure}

Next we want to decouple spin and phonon degrees of freedom. 
The idea is to apply another unitary transformation 
$\bar H  =  \exp(S_2) \tilde H \exp(-S_2)$,
where $S_2$ should be chosen such that all contributions of 
first order in the coupling constant $\bar g$ disappear in the
transformed Hamiltonian. This gives the usual condition for 
a Schrieffer-Wolff transformation~\cite{SW66},
\begin{equation}
  \tilde H_{\rm sp} + [S_2, \tilde H_{\rm s} + \tilde H_{\rm p}] 
  \Big|_{\Delta_\pi=0} \stackrel{!}{=} 0\label{swequ}
\end{equation}
or
\begin{equation}
  S_2 =  {|n\rangle\langle n| \tilde H_{\rm sp}|m \rangle\langle m| 
    \over E_n - E_m}\,,\label{swcond}
\end{equation}
where $|n\rangle$ and $|m\rangle$ denote eigenstates of the 
unperturbed Hamiltonian $H_{\rm s} + H_{\rm p}$ with the 
corresponding eigenenergies $E_n$ and $E_m$.
Since we do not know the exact eigenstates of $H_{\rm s}$, 
equation~(\ref{swcond}) can not be written in a simple form.
One first guess which describes the phonon-part of $S_2$
in a correct way and ensures $S_2$ to be antihermitean is
\begin{eqnarray}
  S_2^{\rm loc} & = & f {\bar g\over\omega_0} 
  \sum_i (b_i^+ - b_i) \sds{i}{i+1}\,,\\
  S_2^{\rm diff} & = & f {\bar g\over\omega_0} 
  \sum_i (b_i^+ - b_i)(\sds{i}{i+1} - \sds{i}{i-1})\,.
\end{eqnarray}
Comparing with (\ref{swcond}) $\omega_0$ gives the appropriate
contribution to the energy denominator, since $H_{\rm sp}$ 
connects only phonon states differing by one in their occupation
number. What is missing is the contribution of spin states
which have overlap via $H_{\rm sp}$.
Therefore, using a numerical procedure we fix the free parameter 
$f$ by the condition, that the amplitude of the state resulting from 
the application of $\tilde H_{\rm sp} + [S_2, \tilde H_{\rm s} + 
\tilde H_{\rm p}]|_{\Delta_\pi=0}$
to the ground state of $H_{\rm s} + H_{\rm p}$ is minimal. 
Figure~\ref{figfopt} illustrates the variation of $f$ with varying 
phonon frequency $\omega_0$.
We find that the general shape of $f=f(\omega_0)$ depends only weakly
on both, system size $N$ and frustration $\alpha$. While 
$f\rightarrow 1$ in the anti-adiabatic frequency range 
($\omega_0\gg J$), the transformation $\exp(S_2)$ vanishes 
completely as the frequency becomes small ($\omega_0\ll J$). 

In contrast to electron-phonon systems with Holstein coupling, 
where a transformation similar to $\exp(S_2)$ completely
removes the electron-phonon interaction term, applying the unitary 
transformation $\exp(S_2)$ to $\tilde H$, we obtain an infinite 
series of terms, which can not be summed up to a simple expression, i.e.
\begin{equation}
  \bar H  =  \sum_k [S_2, \tilde H ]_k/k!\label{transS2}\,,
\end{equation}
where ${[S_2, \tilde H]_k}$ denotes the iterated commutator
$[S_2, \tilde H]_{k+1}  = [S_2, [S_2, \tilde H]_{k}]$ with
$[S_2, \tilde H]_0 = \tilde H$. 

To get an effective spin model we consider only contributions
up to fourth order in~$\bar g$ and average over the phonon subsystem.
The resulting spin Hamiltonian $H_{\rm eff} = \langle\bar H \rangle$
contains long ranged Heisenberg interactions as well as numerous 
four- and six-spin couplings of the form 
$(\sds{i}{j})(\sds{k}{l})\ldots(\sds{m}{n})$. To a good
approximation we can neglect them and obtain
\begin{eqnarray}
  H_{\rm eff}/J & = & J_0 + \sum_i (J_1 + (\textrm{-1})^i
  \delta)\sds{i}{i+1}\label{Heff}\\
  & & + \sum_i \sum_{k=2}^4 J_k \sds{i}{i+k}\,.\nonumber{}
\end{eqnarray}
Note that all phonon dynamics disappeared from the Hamiltonian $H_{\rm eff}$, 
but the effect of the spin-phonon interaction enters through both, 
the static dimerization parameter $\Delta_\pi$ and
the different long range spin interactions.

For the local coupling the corresponding interaction strengths are
\begin{eqnarray}
  \delta^{\rm loc} & = &  - {\Delta_\pi\over J}\left[ (1-f) + 
    {f^2 g^2 Y\over 2}\big(1-{f\over 3}\big)\right]\,,\\
  J_0^{\rm loc} & = & N\Bigg[{\omega_0\over J} \langle b_i^+ b_i\rangle
  + {1\over 4 \lambda} \left({\Delta_\pi\over J}\right)^2\nonumber{}\\
  & & - {3\over 8} \lambda f (1-{f\over 2})
  - {\lambda g^2 Y f^3\over 16}(1-{f\over 4})\Bigg]\,,\\
  J_1^{\rm loc} & = & 1 + \lambda f (1-{f\over 2}) +  
  {f^2 g^2 Y (1-\alpha)\over 2} \\
  & & + {\lambda g^2 Y f^3\over 2}(1-{f\over 4}) 
  + {f^4 g^4 Y^2\over 96}(28-37\alpha)\,,\nonumber\\
  J_2^{\rm loc} & = & \alpha - {f^2 g^2 Y (1-2\alpha)\over 2}\\
  & & - {\lambda g^2 Y f^3\over 4}(1-{f\over 4})
  - f^4 g^4 Y^2{37\over 96}(1-2\alpha)\,,\nonumber\\
  J_3^{\rm loc} & = &  -{f^2 g^2 Y \alpha\over 2} 
  + {f^4 g^4 Y^2\over 96}(9-46\alpha) \,,\\
  J_4^{\rm loc} & = & {9 f^4 g^4 Y^2 \alpha \over 96} \,,
\end{eqnarray}
while for the difference coupling we find
\begin{eqnarray}
  \delta^{\rm diff} & = & 2\ \delta^{\rm loc}\,,\\
  J_0^{\rm diff} & = & N\Bigg[{\omega_0\over J}\langle b_i^+b_i\rangle 
  + {1\over 4 \lambda} \left({\Delta_\pi\over J}\right)^2\nonumber{}\\
  & & - {3\over 4} \lambda f (1-{f\over 2})
  - {3 \lambda g^2 Y f^3\over 16}(1-{f\over 4}) \Bigg]\,,\\
  J_1^{\rm diff} & = & 1 + 2 \lambda f (1-{f\over 2}) +  
  {3 f^2 g^2 Y (1-\alpha)\over 2} \\
  & & + {3 \lambda g^2 Y f^3\over 2}(1-{f\over 4}) 
  + {f^4 g^4 Y^2\over 24}(59-75\alpha)\,,\nonumber\\
  J_2^{\rm diff} & = & \alpha + \lambda f(1-{f\over 2})
  - {f^2 g^2 Y (3-5\alpha)\over 2}\\
  & & - {\lambda g^2 Y f^3\over 12}(1-{f\over 4})
  - {f^4 g^4 Y^2\over 24}(75-124\alpha)\,,\nonumber\\
  J_3^{\rm diff} & = & -f^2 g^2 Y \alpha 
  - {5 \lambda g^2 Y f^3\over 6}(1-{f\over 4})\\
  & & + {f^4 g^4 Y^2\over 48}(32-119\alpha)\,,\nonumber\\
  J_4^{\rm diff} & = & {21 f^4 g^4 Y^2 \alpha \over 48}\,. 
\end{eqnarray}
To point out the relevant model parameters we introduced the dimensionless 
coupling constants $\lambda = \bar g^2/(J\omega_0)$ 
(cf. Refs.~\cite{Pyt74,CF79}) and $g = \bar g/\omega_0$. 
Besides we used
\begin{equation}
  Y  :=  \langle (b_i^+ - b_i)^2 \rangle = \left\{
    \begin{array}{ll}
      -1 & \textrm{if  } T = 0\\
      -\coth({\omega_0\over 2T}) & \textrm{if  } T > 0
    \end{array}\right.\,,
\end{equation}
as a shorthand notation. To compare our result with that of 
Uhrig~\cite{Uhr98}, in $H_{\rm eff}^{\rm diff}$ we have to set
$f=1$, which corresponds to the anti-adiabatic regime,
and $\Delta_\pi=0$. Indeed we recover 
(except for a prefactor $1/2$ which in Ref.~\cite{Uhr98} enters
erroneously going from eq.~(11c) to eq.~(13), and which
consequently is also wrong in Ref.~\cite[eq.~(3-4)]{BMH98})  
all second order terms derived with the flow-equation method, 
supplemented by some new fourth order contributions. 

However, in contrast to Uhrig~\cite{Uhr98},
who proposed to take a thermal average within the phonon subsystem,
we believe, that averaging over the phonon vacuum, i.e. setting
$T=0$, is a more natural choice, especially if we are dealing with
phonon-energies in the anti-adiabatic regime, where the interesting
energy- and temperature-scales are small in respect of $\omega_0$.
Therefore in the following we use $Y=-1$ 
and $\langle b_i^+ b_i\rangle = 0$ exclusively. 

\section{Transition to a gapped phase}

A prominent feature associated with the SP instability is of
course the existence of an energy gap between the ground state
and lowest excitations. Considering, in a first step, the pure
spin model $H_{\rm s}$, it is known that the spectrum is gapless for
the Heisenberg chain with $\alpha = 0$, where the lowest spinon
excitations (triplet and singlet) are degenerate with the ground state
at momenta $q=0$ and $\pi$~\cite{dCP62,Yam69,FT81}. In contrast
the system has a two-fold degenerate ground state and a gap to lowest 
triplet excitations at $\alpha=0.5$, the Majumdar-Ghosh point~\cite{MG69}. 
At some intermediate frustration $\alpha_c$ the model undergoes
a transition from the gapless to the gapped phase, which is of 
Kosterlitz-Thouless type~\cite{BE81,Nij81,Hal82}. 
Using arguments of conformal field theory one can show that 
the lowest singlet and triplet excitations of a finite system
of size $N$ become degenerate at $\alpha_c(N)$, where the dependence
on $N$ is only weak and 
$\alpha_c(N)-\alpha_c(\infty)\sim N^{-2}$~\cite{Car86,EN88,AGS89}.
This was used in Refs.~\cite{ON92,CCE95} to determine $\alpha_c = 0.2411$.

Looking at our effective spin models $H_{\rm eff}$ we find that
the interaction with the optical phonons induces the same kind
of frustrating next-nearest neighbour interaction. Therefore, without
any explicit frustration $\alpha$, the effective frustration
$\alpha_{\rm eff} := J_2/J_1$ due to the phonons can lead to a gap in the 
energy spectrum and to spontaneous dimerization, as was already
discussed in Ref.~\cite{KF87}. This effect is most important
in the anti-adiabatic frequency range.

Another mechanism producing a gap is (static) dimerization, i.e.
an alternation of the nearest neighbour exchange integral. Taking
the adiabatic limit of our effective model, 
$f\rightarrow 0$ and $\delta\rightarrow\Delta_\pi$, 
the ground-state energy of 
$H_{\rm s} + \delta \sum_i (\textrm{-1})^i\sds{i}{i+1}$
is known to deviate from its value at $\delta=0$ like 
$\delta^{4/3}$~\cite{CF79}, while the elastic energy increases
with $\delta^2$.  Therefore for all couplings
$\lambda$ the ground-state energy of $H_{\rm eff}(f=0)$ is
minimal, if $\delta$ is finite. At the same time proportional
to $\delta^{2/3}$ a gap opens in the spectrum.

By taking into account both mechanisms we can now determine the
transition from the gapless phase to the gapped one. As the
SP-system behaves different for the two couplings, we treat them
separately, starting with the {\it local} coupling case.

\subsection{Local coupling}

In a first step we set $\Delta_\pi=0$ and use the level-crossing 
criterium~\cite{ON92,CCE95,BMH98} to calculate the critical line in 
the $\alpha$-$g$-plane for different phonon frequencies $\omega_0$ 
and system sizes $N$. Since $H_{\rm eff}^{\rm loc}$ contains longer 
ranged interactions such as $\sds{i}{i+3}$ and $\sds{i}{i+4}$,
this line slightly deviates from the line $\alpha_{\rm eff} = \alpha_c$,
and we have to calculate it separately. Applying the Lanczos algorithm 
to the effective model we obtain the critical line with high accuracy 
on local workstations ($N\le 20$).
On the other hand we determine the level-crossing in the original 
model~(\ref{model}) by using the methods described in Ref.~\cite{WFK98}. 
In the case $\omega_0/J=0.1$, the latter is complicated for the small 
spin phonon systems we can handle with our Lanczos diagonalization code
on present day parallel computers, 
since the finite size gap to the first triplet excitation is a few times 
larger than $\omega_0$. If $g$ is small enough, each eigenvalues of the 
spin system is accompanied by a ladder of phonon states. Therefore several 
'copies' of the singlet ground state lie within the singlet-triplet 
gap if $\omega_0\ll J$, and the singlet we consider for the level-crossing
is not easy to determine unless we go to very large systems 
(see Ref.~\cite[Tab.II]{BMH98}). 

\begin{figure}[!h]
  \begin{center}
    \epsfig{file= 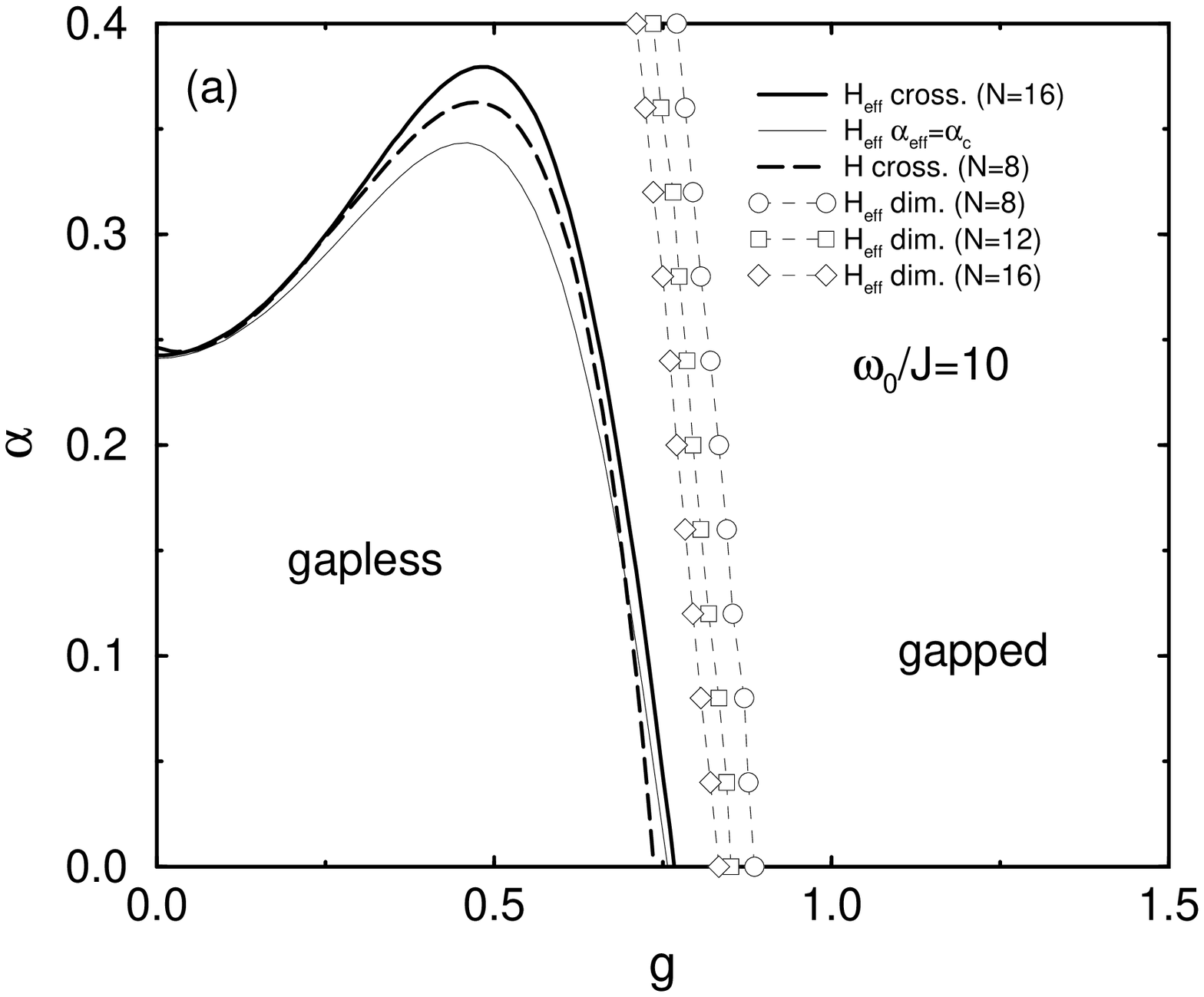, width = 0.9\linewidth}
    \epsfig{file= 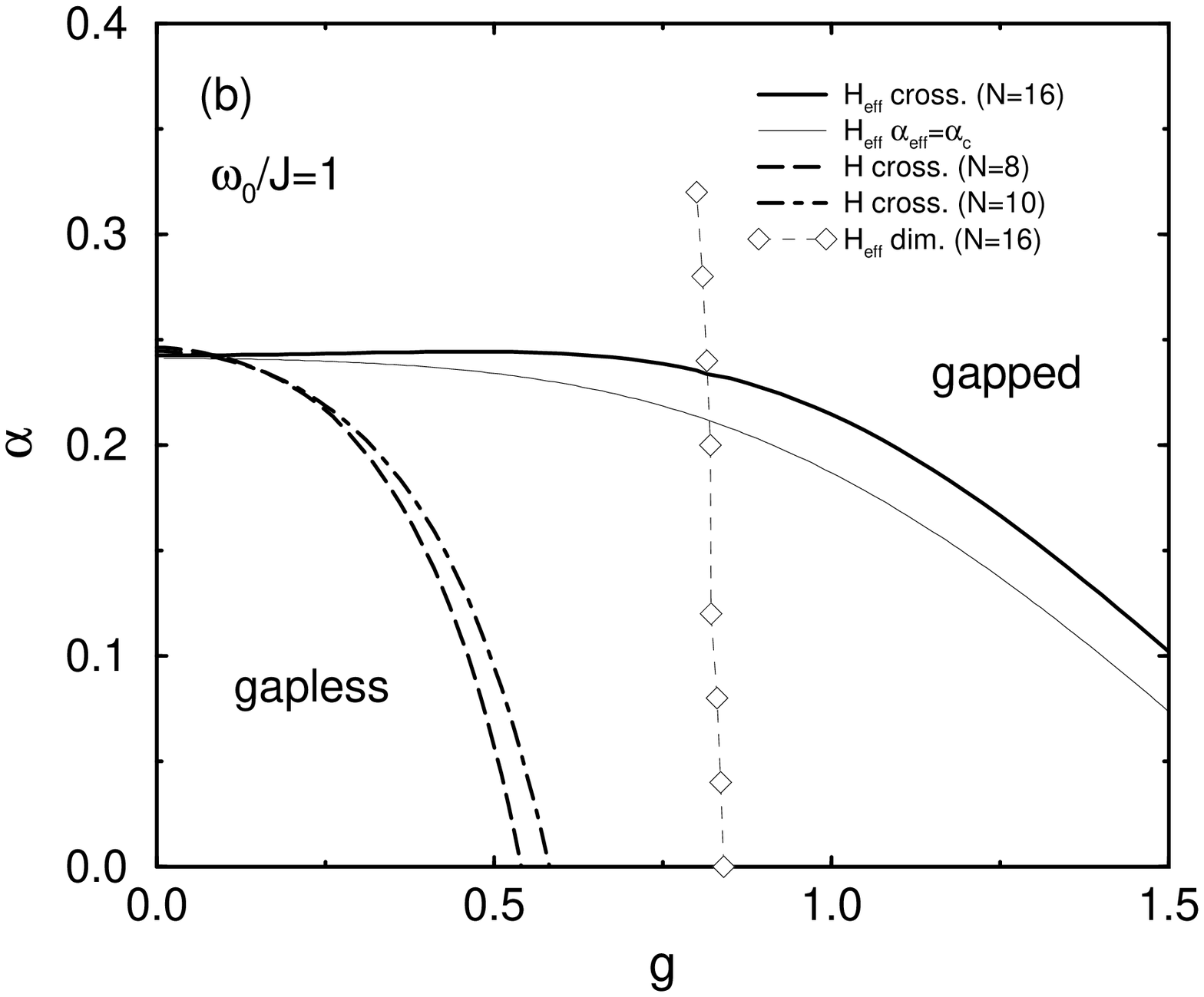,  width = 0.9\linewidth}
    \epsfig{file= 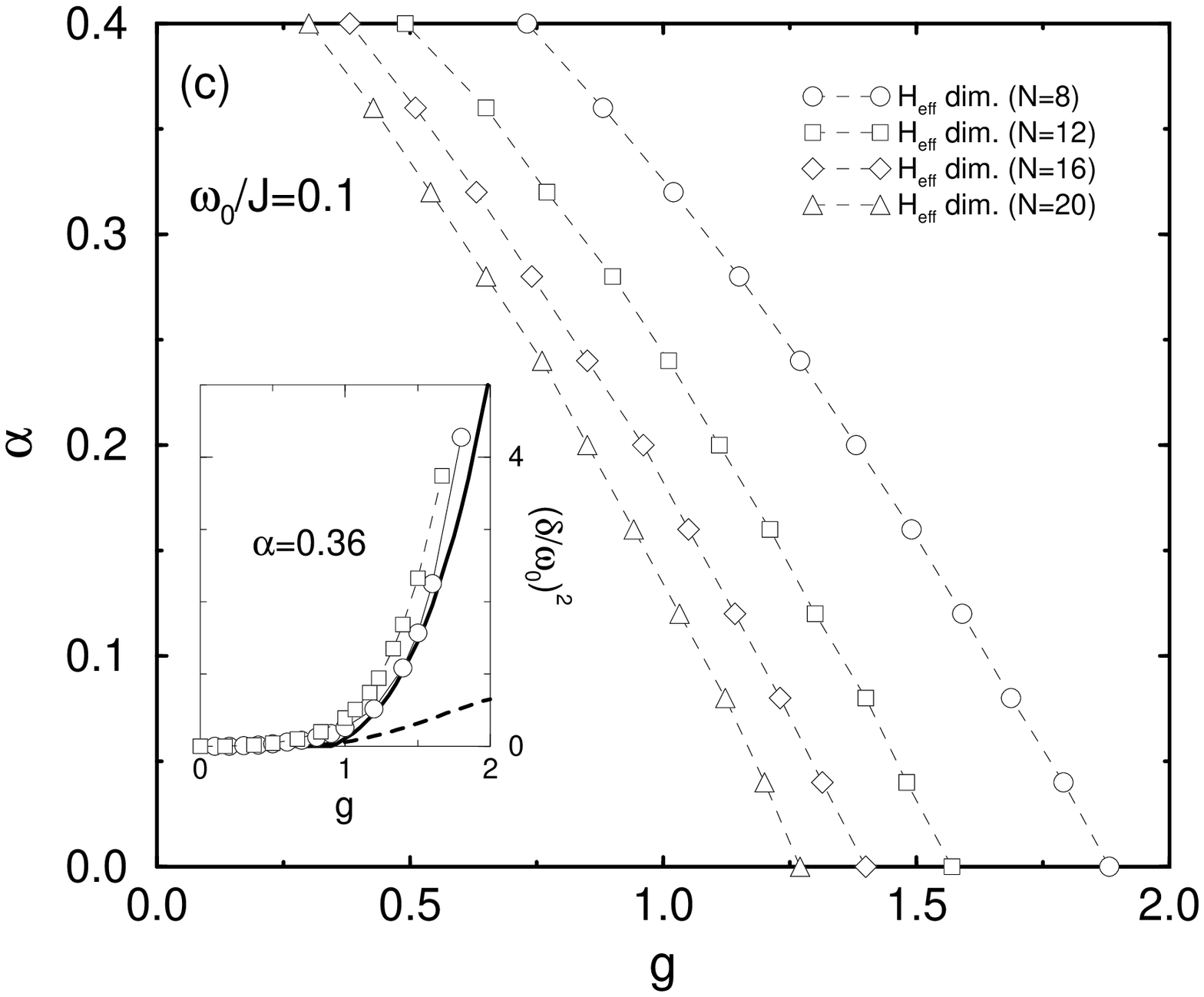, width = 0.9\linewidth}
  \end{center}
  \caption{Singlet-triplet level-crossing (solid lines) and onset of
    dimerization (dashed with symbols) in the effective model
    in comparison to the level-crossing in the original 
    model (bold dashed lines) at $\omega_0/J=10,\ 1$ and $0.1$.
    }\label{figcrloc}
\end{figure}

Figure~\ref{figcrloc}(a) and~\ref{figcrloc}(b) show the critical lines 
in the effective (bold solid) and the original model (bold dashed) as 
well as the lines $\alpha_{\rm eff}=\alpha_c$ (thin solid). As in the
case of the pure spin model, the critical lines depend only weakly on $N$. 
We can therefore compare exact data for the original model and $N=8$ with 
data for the effective model and $N=16$.
While the results differ noticeable for intermediate phonon frequency
$\omega_0\sim 1$, the agreement is excellent in the anti-adiabatic
frequency range $\omega_0\gg J$. 
With increasing $\omega_0$ the critical curve exhibits a remarkable
upturn before crossing the abscissa, i.e. the frustration is  suppressed
for small spin phonon coupling, but overcritical for strong coupling.
It is this feature which makes it necessary to expand~(\ref{transS2})
up to fourth order to approximate $H^{\rm loc}$ in a correct way.
A second order theory is not capable to describe the observed 
critical line.

Another point we can study within our effective model is the 
behaviour of the critical spin phonon coupling $g_c$ at $\alpha=0$ 
in the limit $\omega_0/J \rightarrow \infty$,
i.e. the limit of the crossing point of the critical line and the
abscissa. As the effects of the longer ranged interactions are rather 
small, we can solve the equation $\alpha_{\rm eff}^{\rm loc} = \alpha_c$. 
Setting $f=1$ (compare Fig.~\ref{figfopt}) and $\alpha=0$ we find
\begin{eqnarray}
  g_c^2 & = & {P\over 2Q} + \sqrt{\left({P\over 2Q}\right)^2 
    + {\alpha_c\over Q}}\quad\textrm{with}\\
    P & = & {\omega_0\over J}{\alpha_c\over 2} - {1\over 2}(1+\alpha_c)\,,\\
    Q & = & {\omega_0\over J}{3\over 8} ({1\over 2} + \alpha_c) 
      - ({37\over 96} + {7\over 24}\alpha_c)\,,
\end{eqnarray}
and in the  limit of infinite phonon frequency $g_c$ approaches a 
finite value, 
\begin{equation}
  \lim_{\omega_0/J\rightarrow\infty} g_c = 
  \sqrt{{8 \alpha_c \over 3(1 + 2\alpha_c)}} \approx 0.66\,,
\end{equation}
for the model with local coupling.

\begin{figure}[!htb]
  \epsfig{file= 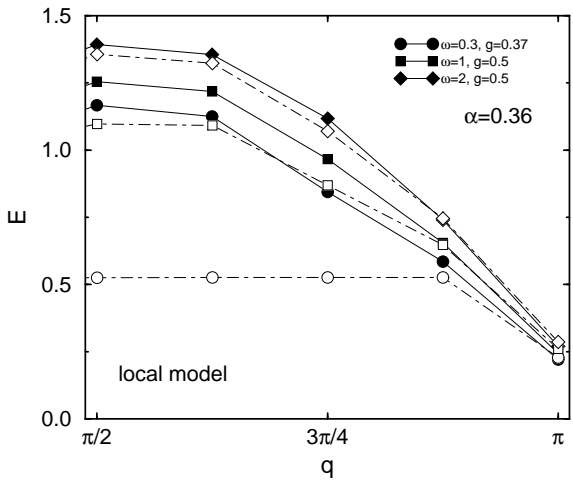, width = \linewidth}
  \caption{Low-lying excitations in the effective (filled symbols) 
    and the original model (open symbols) for different
    frequency and local coupling $g$.}\label{figdispers}
\end{figure}

In the case small phonon frequency, $\omega_0\ll J$, the second
transformation $\exp(S_2)$ looses their importance, and the effective 
frustration due to the spin phonon interaction is replaced by the 
dimerization as the relevant mechanism leading to an energy gap. We 
account for this effect by allowing for a finite $\Delta_\pi$ in our 
approximation. Using the Hellmann-Feynman theorem and numerical
diagonalization of finite spin systems we determine $\Delta_\pi$ 
such that the ground-state energy of $H_{\rm eff}^{\rm loc}$ is 
minimal. Depending on coupling strength $g$ and frustration $\alpha$ 
the system prefers to remain in the undimerized, gapless phase 
($\Delta_\pi=0$) or to develop a non-zero dimerization leading to a gap. 
In Figs.~\ref{figcrloc}(a)~--~(c) we plotted
these transition lines (dashed line with symbols) in addition 
to those obtained by level-crossing. As we already found in our study of the
Holstein model of spinless fermions~\cite{WF98}, for small $\omega_0$
the transition to the dimerized phase depends noticeable on the 
system size $N$ (see Fig.~\ref{figcrloc}(c)), while the finite-size 
dependence is weak in the anti-adiabatic regime 
(cf. Fig.~\ref{figcrloc}(a)). In addition, for $\omega_0/J = 10$ 
the transition is consistent with the critical line determined 
via level-crossing. 

\begin{figure}[htb]
  \epsfig{file= 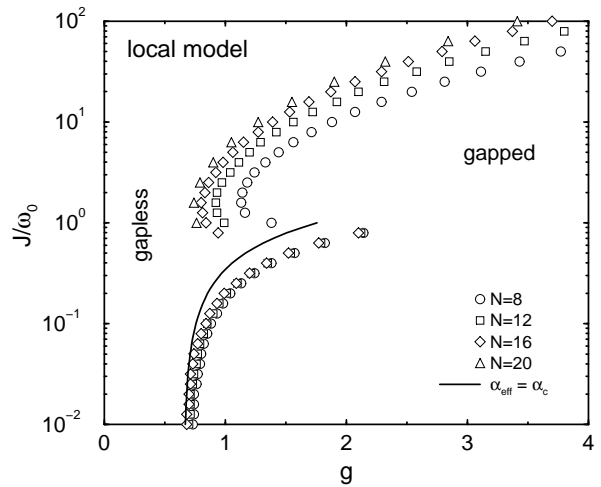, width = \linewidth}
  \caption{Critical coupling $g_c$ versus frequency for 
    $H_{\rm eff}^{\rm loc}$ with $\alpha=0$.}\label{figphdialoc}
\end{figure}

To compare properties of the original and the effective model
also in the case of small phonon frequency, we consider the 
dimerization. As a quantity which corresponds to $\delta^{\rm loc/diff}$ 
we take the static (lattice) structure factor~\cite{WFK98,BFK98},
\begin{equation}
  \delta^2 = {\bar g^2\over N^2 }
  \sum_{j,k} \langle u_j u_k \rangle e^{{\rm i}\pi (j-k)}\,,
\end{equation}
calculated in the ground state of~(\ref{model}), where $u_j = b_j^+ + b_j$. 
The inset of Fig.~\ref{figcrloc}(c) displays $(\delta/\omega_0)^2$ for 
$\alpha=0.36$, $\omega_0/J=0.1$ (solid line), $0.316$ (dashed line), 
and $N=8$. The results for the effective model (bold lines) and the original 
model (thin lines with symbols) agree rather well, especially 
for $\omega_0/J = 0.1$.

Another feature we can compare is the dispersion of low-lying
excitations. Figure~\ref{figdispers} shows the energy
of the lowest triplet excitations, calculated exactly and within our
approximation. Clearly for $\omega_0/J=0.3$ the correct 
dispersion is flattened at momenta near $q=\pi/2$. This results
from the energy of the dispersionless phonons, which is added 
to the lowest triplet at $q=\pi$.  Of course our effective model
does not contain these low-lying phonon excitations.
However, as soon as $\omega_0\gtrsim J$ the
lowest excitations are due to renormalized spin interactions and 
well approximated by the effective model.

To collect the results of this subsection we show in
Fig.~\ref{figphdialoc} the critical coupling $g_c(\alpha=0)$ 
over a wide range of phonon frequencies, using both criteria
for the phase transition. Symbols stand for the onset 
of dimerization, while the bold line corresponds to 
$\alpha_{\rm eff}=\alpha_c$. As expected, we find that our approximation
is somewhat unreliable for intermediate phonon frequency.
The singularity of $g_c$ at $\omega_0/J = 1$ is a 
manifestation of this deficiency. The correct critical 
line will connect adiabatic and anti-adiabatic behaviour in 
a continuous way (compare also next subsection and~\cite{BMH98}).

\subsection{Difference coupling}

\begin{figure}[!htb]
  \begin{center}
    \epsfig{file= 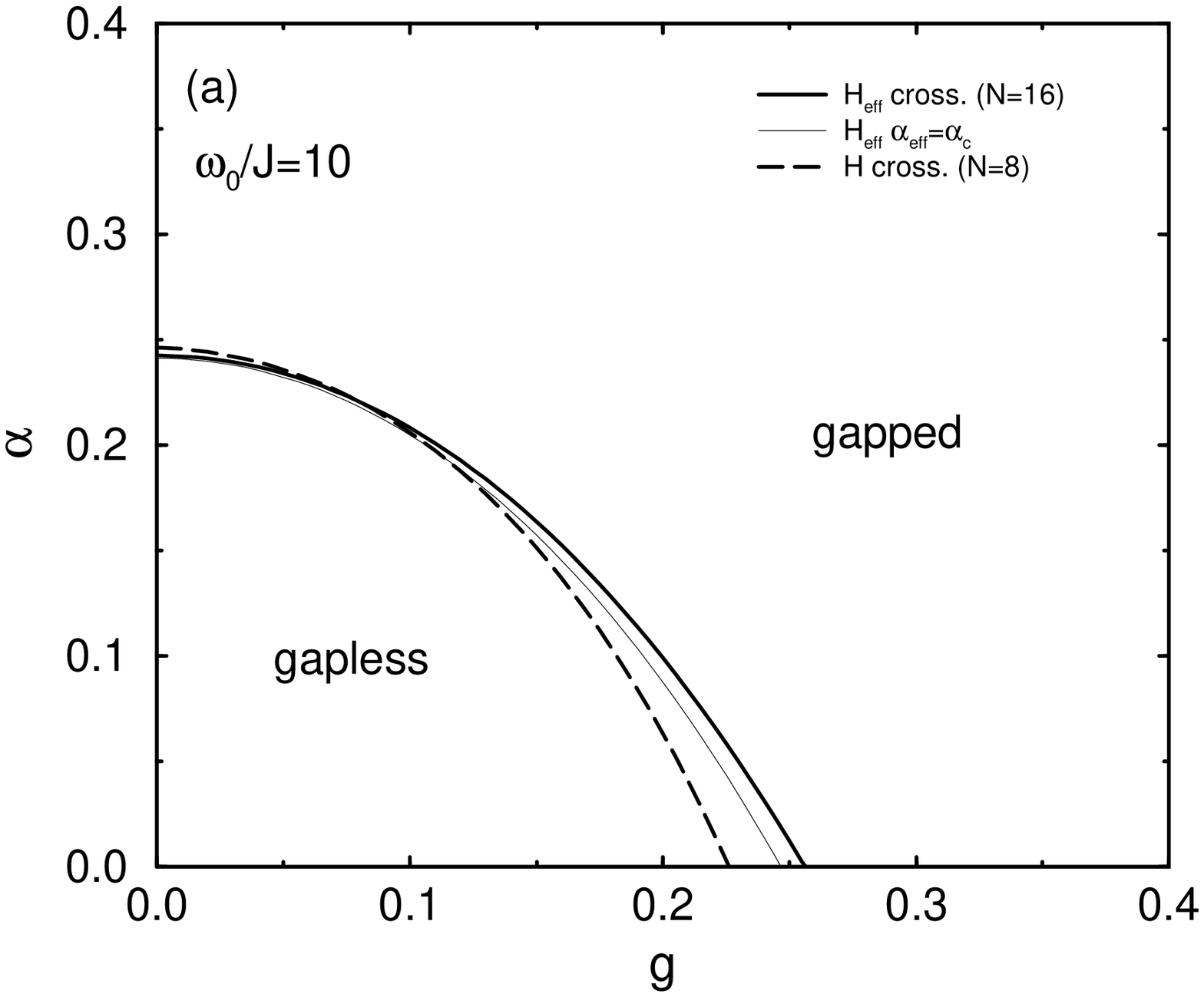, width = 0.9\linewidth}
    \epsfig{file= 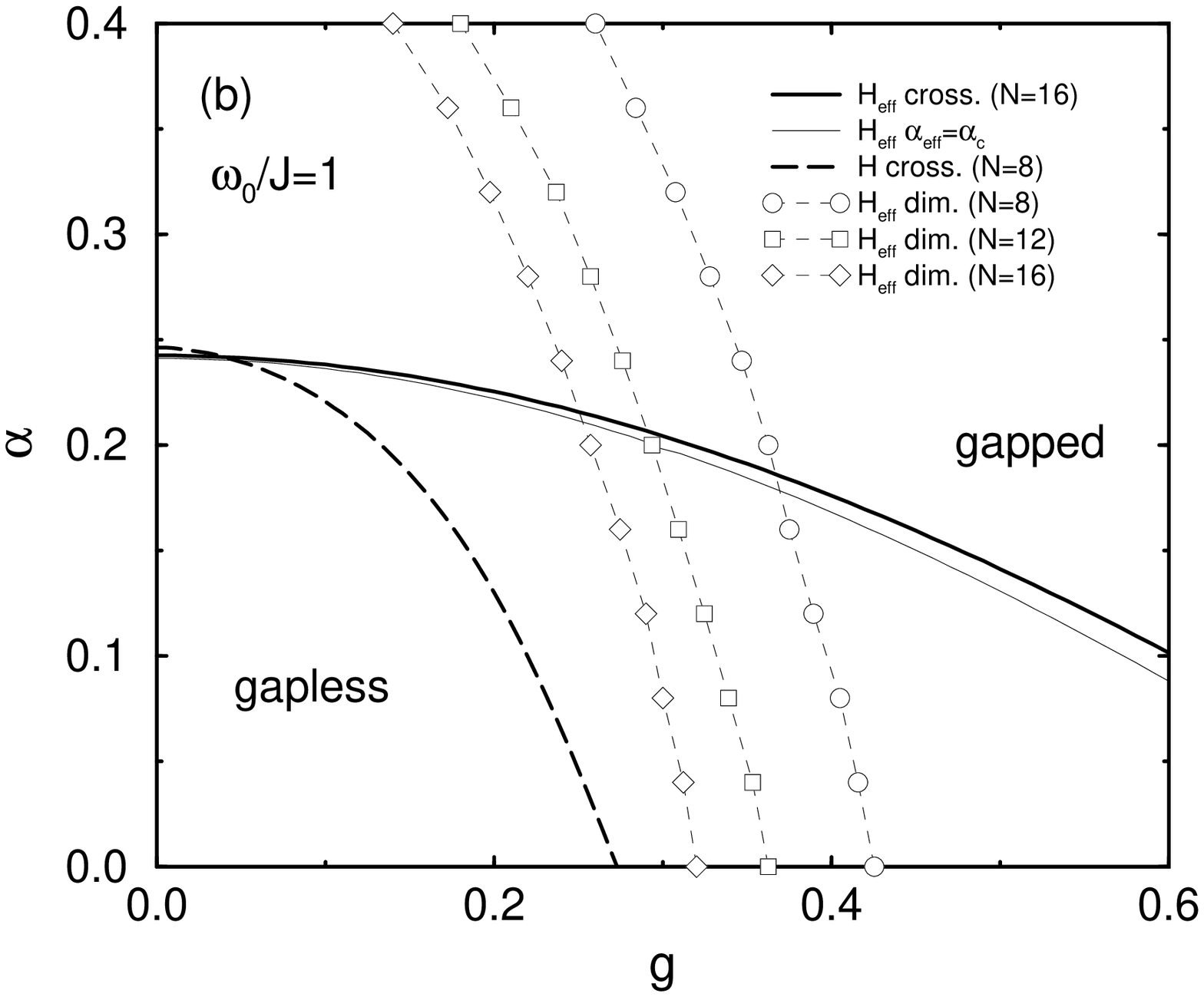,  width = 0.9\linewidth}
    \epsfig{file= 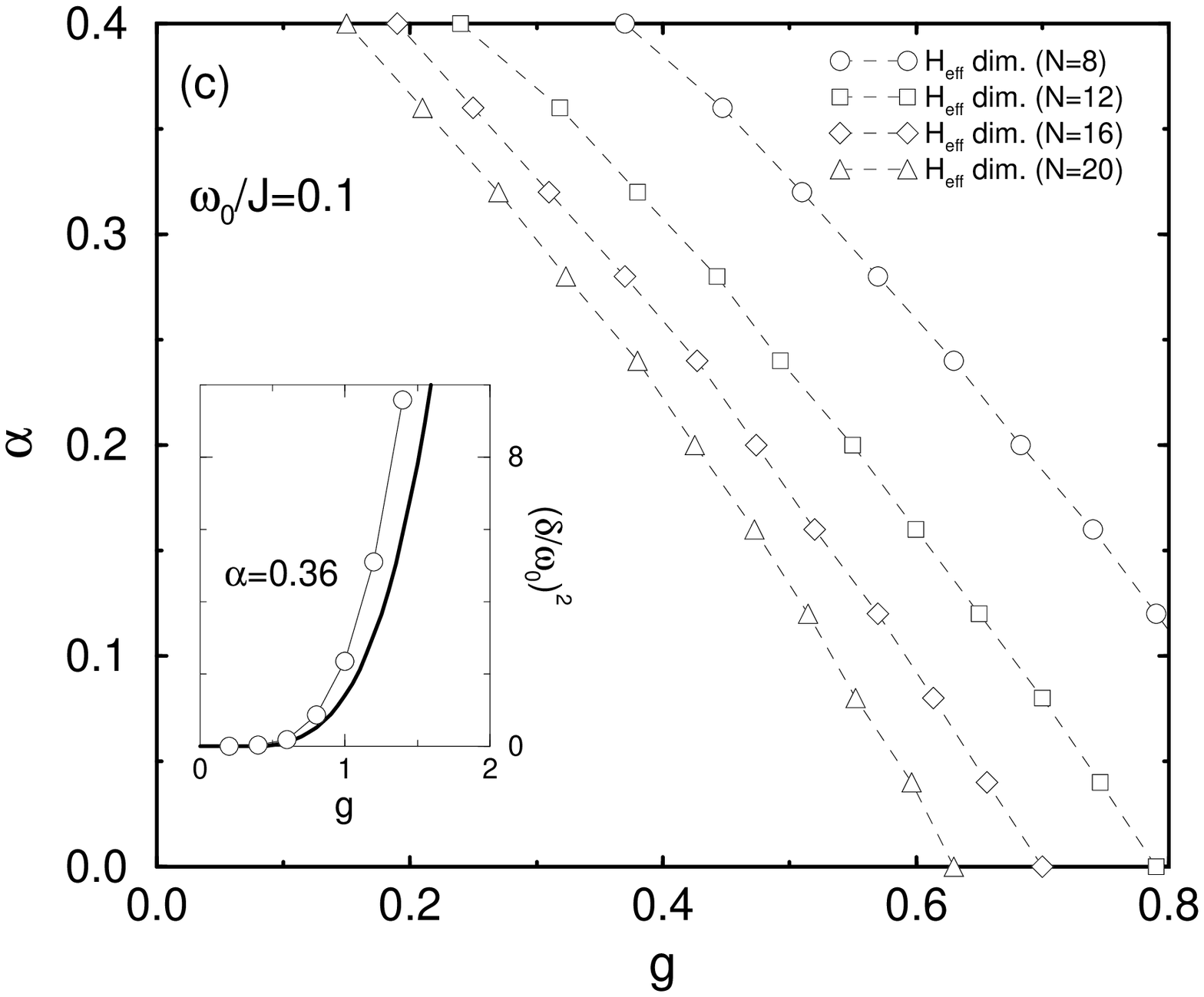, width = 0.9\linewidth}
  \end{center}
  \caption{Level-crossing (solid lines) and onset of
    dimerization (dashed with symbols) in the effective model
    in comparison to the level-crossing in the original 
    model (bold dashed lines) at $\omega_0/J=10,\ 1$ and $0.1$.
    }\label{figcrdif}
\end{figure}

The procedure to determine the phase transition in the 
SP-system with difference coupling
is the same as described before. In the anti-adiabatic regime we 
set $\Delta_\pi=0$ and calculate the position of the crossing of 
the first triplet and the first singlet excitation for both, the
original and the effective model. The results for 
$\omega_0/J=10$ and $1$ are shown in Figs.~\ref{figcrdif}(a)
and~\ref{figcrdif}(b) respectively.
In contrast to the local coupling the structure of the 
critical line for high phonon frequency ($\omega_0/J=10$)
is much simpler. It appears that one would get the 
same shape also for a second order theory. However, to enlarge
the application area of our approximation taking into 
account higher order contributions is still appropriate. As before,
the agreement between the original and the effective model
is excellent in the anti-adiabatic regime, while the deviations
increase with approaching intermediate frequencies.

Calculating the behaviour of the critical coupling, $g_c(\alpha=0)$
in the limit of infinite phonon frequency,
$\omega_0/J\rightarrow\infty$, we now find
\begin{eqnarray}
  g_c^2 & = & -{P\over 2Q} + \sqrt{\left({P\over 2Q}\right)^2 
    + {\alpha_c\over Q}}\quad\textrm{with}\\
    P & = & {\omega_0\over J}({1\over 2} - \alpha_c) 
    + {3\over 2}(1+\alpha_c)\,,\\
    Q & = & {\omega_0\over J}({1\over 16} + {9\over 8}\alpha_c) 
      - ({25\over 8} + {59\over 24}\alpha_c)\,,
\end{eqnarray}
and, different to the local coupling case, $g_c$ tends to zero,
\begin{equation}
  \lim_{\omega_0/J\rightarrow\infty} g_c = 0\,.
\end{equation}
While the $q=0$ and the $q=\pi$ phonon mode compete in the 
case of local spin phonon coupling, allowing for a stable
gapless phase up to a critical $g$, there is no interaction
with the $q=0$ mode in $H_{\rm sp}^{\rm diff}$. Therefore the 
$q=\pi$ mode induces long ranged exchange more efficiently,
leading to a vanishing $g_c$ for $\omega_0/J\rightarrow\infty$.

For small phonon frequencies $\omega_0\ll J$ again we determine the optimal
dimerization $\Delta_\pi$ and the critical line beyond which
$\Delta_\pi$ starts to be non-zero. The results are shown in 
Figs.~\ref{figcrdif}(b) and~\ref{figcrdif}(c). For large
phonon frequency we find an unstable behaviour of $\Delta_\pi$,
it is finite for some $g$, but vanishes before growing to
substantial values again. We therefore make no attempt 
to fix the onset of $\Delta_\pi\ne 0$ for $\omega_0/J\gtrsim 1$.

As in the case of local coupling the dimerization 
in the original and the effective model agrees well
for $\omega_0/J=0.1$ (inset Fig.~\ref{figcrdif}(c)).

\begin{figure}[!htb]
  \epsfig{file= 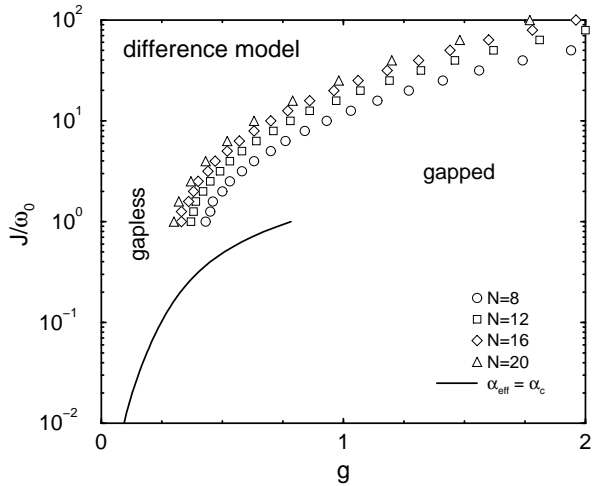, width = \linewidth}
  \caption{Critical coupling $g_c$ versus frequency for 
    $H_{\rm eff}^{\rm diff}$ with $\alpha=0$.}\label{figphdiadif}
\end{figure}

Finally in Fig.~\ref{figphdiadif} we combine the above results to
obtain the phase diagram in the $J/\omega_0$-$g$ plane.
Except for $\omega_0/J\rightarrow\infty$, where $g_c\rightarrow 0$, 
the behaviour is  similar to the case of local coupling. Again 
at $\omega_0/J\sim 1$ the effective model gives a singularity in 
$g_c$, while the correct result should be continuous. Comparing the 
transition line with the very recent DMRG data of Bursill 
et~al.~\cite{BMH98}, we find very good agreement in the anti-adiabatic 
regime (the phase transition does not change, going from the second to the 
fourth order effective theory). For small phonon frequency 
$\omega_0/J\rightarrow 0$ the results of
Bursill~et~al. suggest a finite limit for $g_c/\omega_0$, while
our model gives an increasing ratio of $g_c/\omega_0$.
However, as the finite size effects are large for $\omega_0\ll J$,
determining the correct value of $g_c$ is a delicate procedure
and the exact $g_c$ might be much smaller than depicted in 
Fig.~\ref{figphdiadif}.

\section{Ground-state phonon distribution}

In the course of exact diagonalizations of the
phonon dynamical model~(\ref{model}) we observed another
interesting feature distinguishing the two mechanisms
of spin phonon interaction. Turning our attention to the 
phonon distribution in the ground state of~(\ref{model}),
\begin{figure}[!htb]
  \epsfig{file= 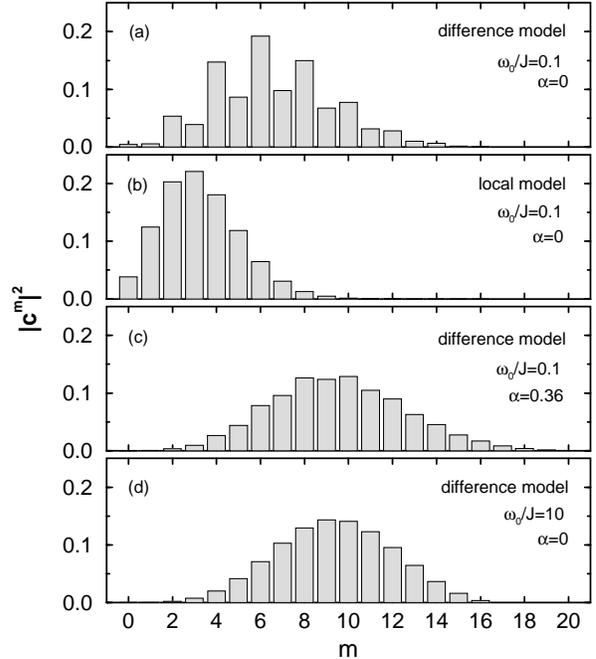, width = \linewidth}
  \caption{Even/odd-phonon distribution in the ground state of the
    model with difference coupling ($g=1.4$, $\omega_0/J=0.1$) and
    mechanisms that suppress the effect: local coupling, frustration
    and anti-adiabaticity.
    }\label{figphonon}
\end{figure}
we find for the model with {\it difference} coupling in the
adiabatic regime, that the system prefers states with {\it even} 
phonon occupation numbers (cf. Fig.~\ref{figphonon}(a)). 
This behaviour reminds of a simple two-level system, more precisely 
the Rabi (pseudo-Jahn-Teller) Hamiltonian~\cite{PW84}
${H = \Delta\sigma_z+\bar g (b^+ + b)\sigma_x +\omega_0 b^+ b\ \sigma_0 }$,
where an interaction with a bosonic degree of freedom connects the levels.
Solving this model in the adiabatic strong-coupling case, 
${\omega_0/\Delta\ll 1}$ and ${g=\bar g/\omega_0 > 1}$, we obtain a similar 
even/odd oscillation in the ground-state phonon distribution. One can 
understand this effect within standard perturbation theory. Starting with the 
two levels $\pm \Delta$ and the corresponding eigenstates $1\choose 0$  
and $0\choose 1$, it is obvious, that adding only {\it one} phonon 
requires an extra amount of $2\Delta$ exchange energy,
because the system is excited to the upper level. 
Therefore the system prefers an even number of phonons for 
the ground-state wave function. 
In the anti-adiabatic case, $\omega_0\gg \Delta$, to a good 
approximation we can omit the $\Delta\sigma_z$ term and solve the 
model by unitary transformation, leading to a Poisson distribution 
of the phonons (coherent state). 

In view of the SP system with difference coupling, the $q=\pi$ phonon
is the most relevant (in fact in the ground state almost all phonons occupy 
this mode), and adding one phonon changes the
momentum of the spin system by $\pi$. Consequently the gap between 
the (singlet) ground state at $q=0\ (N=4k,\ k\in\mathbb{N})$ or 
$\pi\ (N=4k+2)$ and the lowest singlet state at $q=\pi$, or $0$ 
respectively, plays the role of $2\Delta$ in the two-level system.
As this singlet-singlet gap is only due to the finite size of
the considered spin chain, we believe that the even/odd oscillations
will disappear in the thermodynamic limit. The reduction (disappearance) 
of the even/odd imbalance resulting from a finite frustration with a 
much smaller singlet-singlet gap (zero at $\alpha=0.5$), cf. 
Fig.~\ref{figphonon}(c), can be taken as a first indication.
Of course we observe a smooth phonon distribution in the anti-adiabatic 
case (Fig.~\ref{figphonon}(d)).

The model with local spin phonon coupling exhibits the usual Poisson
distribution for all frequencies, since the interaction is 
due to the $q=\pi$ {\it and} the $q=0$ phonon mode 
(Fig.~\ref{figphonon}(b)).

\section{Conclusion}

In summary, we have studied the spin-Peierls instability of a 
frustrated Heisenberg spin chain coupled to optical phonons of 
energy $\omega_0$. Using the concept of unitary transformations 
we derive effective spin Hamiltonians, which cover the 
spin phonon interaction by two mechanisms, static 
dimerization $\Delta_\pi$ and long ranged exchange couplings. 
Both can lead to an energy  gap between the ground state and 
lowest excitations, which is related to a Peierls instability of the 
spin system. In the anti-adiabatic phonon frequency range, $\omega_0\gg J$, 
we verify and extend the Hamiltonian obtained recently with
the flow-equation method~\cite{Uhr98}, while we recover the usual
static SP model in the limit $\omega_0/J\rightarrow 0$.

To determine the transition to the gapped phase in the case of 
large phonon frequency we use the level-crossing criterium, which
proved to be very accurate for similar models~\cite{ON92,CCE95,BMH98}.
For the two types of spin phonon coupling ({\it local}: $u_i\ \sds{i}{i+1}$, 
and {\it difference}: $(u_i - u_{i+1})\ \sds{i}{i+1}$) we 
consider here, the results of our effective models agree very well 
with data from exact diagonalization of the original, phonon dynamical 
model and with recent DMRG data~\cite{BMH98} (difference 
coupling only). In the case of local coupling two phonon modes 
($q=0$ and $\pi$) compete and allow for a gapless phase to exist
in a wider parameter range. Furthermore we observe a non-monotonous 
behaviour of the phase transition line as a function of spin-phonon
interaction and frustration $\alpha$.
With difference coupling spins and phonons interact almost only through 
the $\pi$-mode, which is able to dimerize the system most efficiently.

For phonon frequencies $\omega_0\lesssim J$, we determine
the phase transition by means of the static dimerization $\Delta_\pi$, which
changes from zero at small spin phonon coupling to a finite value
beyond a critical coupling. 

At intermediate frequency the situation remains unsatisfactory as the
critical coupling behaves discontinuously. Here numerical methods, including 
the full phonon dynamics, still provide the only reliable tool to study
the transition.
Nevertheless the proposed effective models help to understand the physical 
mechanisms leading to spontaneous dimerization of the interacting 
spin-phonon system.

\section{ Acknowledgement}

We thank J.~Schliemann for valuable discussions. Some computations 
were done at LRZ M\"unchen, HLRZ J\"ulich, HLRS Stuttgart, and GMD Bonn.
H.F. acknowledges financial support from the Graduiertenkolleg 
'Nichtlineare Spektroskopie und Dynamik' at the University of Bayreuth.

\end{document}